\documentclass[12pt]{article}
\input epsf.sty
\def\be{\begin{equation}}
\def\ee{\end{equation}}
\def\ba{\begin{align}}
\def\ea{\end{align}}

\def\p{\partial}

\def\ops[#1]{\p_{#1} e^{-2\phi}}

\def\eq[#1]{equation (\ref {eq:#1})}
\def\Eq[#1]{Equation (\ref {eq:#1})}
\def\e[#1]{\ref {eq:#1}}
\def\at[#1]{| _{#1}}

\let\oldpercent\%\renewcommand{\%}{\scalebox{0.85}{\oldpercent}}

\begin{document}

\baselineskip=18pt

\begin{center}
{\Large \bf{Negative single-trace $T\bar T$ holography\\versus\\de Sitter holography}}

\vspace{10mm}

\textit{Amit Giveon}
\break

Racah Institute of Physics, The Hebrew University \\
Jerusalem, 91904, Israel

\end{center}


\vspace{10mm}

\begin{abstract}

In single-trace $T\bar T$ holography, for the negative sign of the deformation,
the geometries dual to high energy states consist of a singular UV wall that separates
black strings in the IR from an asymptotically linear dilaton spacetime.
In the hologram, the black strings amount to states
in a symmetric product of $T\bar T$ deformed $CFT_2$,
whose energy and momentum are split equally among the different blocks.
The properties of this holographic duality have intriguing similarities
with the proposed principles of de Sitter holography.
In this note, the analogy between the two is presented.
In particular, as the horizon location approaches the UV wall,
one obtains a state of maximum finite entropy and infinite temperature.
On the other hand, the bulk temperature of the geometry obtained in the limit is small.
These properties are identical to those proposed in empty de~Sitter holography.

\end{abstract}
\vspace{10mm}


Consider supersting theory on the deformed BTZ black holes background,
formed in the near $k$ NS5 branes on $S^1\times T^4$ with $p$ (negative)
fundamental strings wrapping $S^1$, whose radius we denote by $R$, and which carry momentum number $n$ on this circle.
There are two possible signs of the deformation parameter, denoted by $\lambda\equiv\pm\alpha'/R^2$
in~\cite{Chakraborty:2020swe,Chakraborty:2023mzc,Chakraborty:2023zdd}; here we consider the $\lambda<0$ case
(which amounts to $p$ negative strings,~\cite{Chakraborty:2020swe}):
\be\label{l}
\lambda\equiv -{\alpha'\over R^2}<0~.
\ee
The background thus obtained describes a rotating black string~\footnote{Which
we may regard (upon KK reduction) as a two-dimensional black hole with winding and momentum fundamental string charges, $p$ and $n$,
respectively, as in~\cite{Giveon:2005mi} with $(q_L,q_R)=\left({n\over R}+{pR\over\alpha'},{n\over R}-{pR\over\alpha'}\right)$
(note that a subset of the backgrounds in appendix C of~\cite{Giveon:2005mi} are identical to those below;
we shall not present the coordinate transformation between the two here).} in a three-dimensional spacetime,
denoted by ${\cal M}_3^-$ in~\cite{Chakraborty:2020swe};~\footnote{Empty ${\cal M}_3^-$
is a current-current deformation of massless BTZ in a negative direction,~\cite{Giveon:2017nie};
from the point of view of the boundary at infinity, this background can be thought of as a vacuum of Little String Theory
which contains a large number of negative strings,~\cite{Chakraborty:2020swe}.}
its metric, $B$-field and dilaton appear in equations (3.10) with (3.11)) of~\cite{Chakraborty:2023zdd}:
\be\label{bs}
ds^2=-{N^2\over 1-{\rho^2\over R^2}}d\tau^2+{d\rho^2\over N^2}+{\rho^2\over 1-{\rho^2\over R^2}}\left(d\varphi-N_\varphi d\tau\right)^2~,
\ee
\be\label{btautheta}
B_{\tau\varphi}={\rho^2\over r_5}\sqrt{\left(1-{\rho_-^2\over R^2}\right)\left(1-{\rho_+^2\over R^2}\right)}{1\over 1-{\rho^2\over R^2}}~,
\ee
\be\label{dilaton}
e^{2\Phi}={kv\over p}\sqrt{\left(1-{\rho_-^2\over R^2}\right)\left(1-{\rho_+^2\over R^2}\right)}{1\over 1-{\rho^2\over R^2}}~,
\ee
where
\be\label{nn}
N^2={(\rho^2-\rho_+^2)(\rho^2-\rho_-^2)\over r_5^2\rho^2}~,\qquad N_\varphi={\rho_+\rho_-\over r_5\rho^2}~,
\ee
\be\label{vr5}
r_5\equiv\sqrt{k\alpha'}~,\qquad v\equiv {\rm Volume}(T^4)/\left(2\pi\sqrt{\alpha'}\right)^4~,
\ee
and
\be\label{rrr}
\rho_-\leq\rho_+\leq R
\ee
are the locations of the inner horizon and outer horizon of the black string and the singular wall in ${\cal M}_3^-$,
respectively.~\footnote{See~\cite{Giveon:2017nie,Chakraborty:2020swe,Chakraborty:2023zdd} for more details.}

The angular momentum, $n$, of the black string,~(\ref{bs})--(\ref{vr5}),~\footnote{Which
originates from the momentum $n$ carried by the $p$ negative strings that form it.} is,~\cite{Chakraborty:2023zdd},
\be\label{n}
n={p\over\alpha'}{\rho_-\rho_+\over\sqrt{\left(1-{\rho_-^2\over R^2}\right)\left(1-{\rho_+^2\over R^2}\right)}}~.
\ee
Below, we inspect the properties of the one-parameter family of the backgrounds above for certain $R,k,p$ and $n$.
It can be parameterized e.g. by the location of the event horizon,~$\rho_+$.
Along this one-parameter family, the entropy and temperature are,~\cite{Chakraborty:2023zdd},
\be\label{entropy}
S={2\pi kp\over r_5}{\rho_+\over\sqrt{1-{\rho_-^2\over R^2}}}~,
\ee
and
\be\label{t}
T={\rho_+\over 2\pi r_5 R}{1-{\rho_-^2\over\rho_+^2}\over\sqrt{1-{\rho_+^2\over R^2}}}~,
\ee
respectively, where $\rho_-$ is related to $\rho_+$ via equation~(\ref{n}).
The energy of the black hole is,~\cite{Chakraborty:2023zdd},
\be\label{e}
E={p\over\lambda R}\left\{-1+\left[1-\left(\lambda n\over p\right)^2\right]\sqrt{\left(1-{\rho_-^2\over R^2}\right)\left(1-{\rho_+^2\over R^2}\right)}\right\}~,
\ee
where $\lambda$ is given in~(\ref{l}).

A few comments are in order:
\begin{enumerate}
\item
As usual, the interior of the black hole, $\rho_-<\rho<\rho_+$, is time dependent (with time being $\rho$),
while the exterior of the black hole, $\rho>\rho_+$, is time independent,
with $\tau$ ($\varphi$) being a timelike direction when $\rho<R$ ($\rho>R$), instead of $\rho$.
We shall refer to the regime between the event horizon and the UV wall, $\rho_+<\rho<R$, as the static patch.~\footnote{The
other static regime, $\rho>R$, a.k.a. the asymptotically linear dilaton regime beyond the UV wall,
plays an important role in single-trace $T\bar T$ holography,~\cite{Chakraborty:2020swe,Chakraborty:2023zdd};
we shall return to this point below.}
\item
As one varies the location of the horizon $\rho_+$ (while keeping $R$ and $n$ fixed,~(\ref{n})), one interpolates between
empty $\mathcal{M}_3^- $ (when $\rho_\pm=0$) and the case of a maximal black hole size,
$\rho_+=R$ (in which case $\rho_-=0$,~(\ref{n})) in $\mathcal{M}_3^- $.~\footnote{We
shall discuss the $\rho_+\to R$ limit in detail below.}
\item
The energy is minimal, $E_{min}=0$, for empty $\mathcal{M}_3^- $
(for any deformation parameter,~(\ref{l}), of the massless BTZ black hole),
and is maximal,
\be\label{emax}
E_{max}={pR\over\alpha'}~,
\ee
when the black hole size is maximal.
\item
The entropy is minimal, $S_{min}=0$, for empty $\mathcal{M}_3^- $, and is maximal,
\be\label{smax}
S_{max}=\beta_{bh}E_{max}~,\qquad\beta_{bh}\equiv 2\pi r_5~,
\ee
when $\rho_+=R$.~\footnote{Note that $\beta_{bh}$ in~(\ref{smax}) is {\it not} the inverse $T$ in~(\ref{t}); instead,
it is the inverse temperature of an $SL(2)_k/U(1)$ black hole, that will show up in the $\rho_+\to R$ limit (see below).}
\item
The temperature diverges when the black hole size is maximal,
\be\label{tmax}
T\to\infty\quad  {\rm when}\quad \rho_+\to R~.
\ee
\end{enumerate}

The $\rho_+\to R$ limit is intriguing; we pause to inspect it next.

When $\rho_+\to R$, $\rho_-\to 0$ for {\it any} finite $n$,~(\ref{n}),
and for $\rho\neq 0,R$, the metric, $B$-field and dilaton,~(\ref{bs}),~(\ref{btautheta}) and~(\ref{dilaton}), respectively, approach
\be\label{limit}
ds^2\to{R^2\over r_5^2}d\tau^2-{r_5^2\over R^2-\rho^2}d\rho^2+{R^2\rho^2\over R^2-\rho^2}d\varphi^2~,\quad
B\to 0~,\quad\Phi\to-\infty-\log\sqrt{R^2-\rho^2}~,
\ee
in the limit.
Note that $B=0$ for {\it any} $\rho$, in the limit, and $e^{2\Phi}\to 0$ for any $\rho\neq R$.
Moreover, any point in the static patch between $R$ and $\rho_+$ has shrunk to the singularity,
a.k.a. $\rho\to R$  for any $\rho_+<\rho<R$,
and $N_\varphi\to 0$ for any $\rho\neq 0$,~(\ref{nn}), in the limit.

In the time-dependent regime $0<\rho<R$ of~(\ref{limit}), a change of coordinate,~\footnote{We
take $\alpha'\equiv l_s^2=1$ here and below, when it ain't present.}
\be\label{rhophi}
\rho=R\cos\phi~,
\ee
gives
\be\label{gives}
ds^2={R^2\over r_5^2}d\tau^2-r_5^2d\phi^2+R^2\cot^2\phi d\varphi^2~,\quad \Phi=-\infty-\log(\sin\phi)~,
\ee
which is a $2d$ cosmology with time $0<\phi<\pi/2$ and compact space $\varphi$, times a non-compact specelike direction $\tau$,
and a vanishing string coupling for any $\phi>0$;
it is singular at $\phi=0$ (a.k.a. $\rho=R$) and has a horizon at $\phi=\pi/2$ (a.k.a. $\rho=0$).

In the regime $\rho>R$ of~(\ref{limit}), a change of coordinate,
\be\label{rhophib}
\rho=R\cosh\phi~,
\ee
gives
\be\label{givesb}
ds^2={R^2\over r_5^2}d\tau^2+r_5^2d\phi^2-R^2\coth^2\phi d\varphi^2~,\quad \Phi=-\infty\pm{i\pi\over 2}-\log(\sinh\phi)~,
\ee
which is the geometry of the regime beyond the singularity of an $SL(2)_k/U(1)$ black hole,~\footnote{For a review,
see e.g. \cite{Giveon:1994fu}; the relation of the level $k$ of $SL(2)_k$ and the $r_5$ in~(\ref{givesb}) is that in~(\ref{vr5}).}
with radial direction $\phi>0$ and {\it compact time} $\varphi$,
times a spacelike direction $\tau$.
The $2d$ cosmological geometry above is thus the interior of an $SL(2)_k/U(1)$ black hole with compact time $\varphi\sim\varphi+2\pi$.

The exterior of the $SL(2)_k/U(1)$ black hole amounts to a continuation of
the deformed BTZ geometry~(\ref{bs})--(\ref{vr5}) beyond its singularity (at $\rho=0$).
Concretely, in the limit~(\ref{limit}), the continuation
\be\label{rhophic}
\rho=iR\sinh\phi
\ee
gives
\be\label{givesc}
ds^2={R^2\over r_5^2}d\tau^2+r_5^2d\phi^2-R^2\tanh^2\phi d\varphi^2~,\quad \Phi=-\infty-\log(\cosh\phi)~.
\ee
The horizon of the $SL(2)_k/U(1)$ black hole with compact time in the $(\varphi,\phi)$ directions is at $\rho=0$,
a.k.a. at $\phi=0$ in~(\ref{givesc}),
and its inverse temperature $\beta_{bh}$ is the one presented in~(\ref{smax});
we shall refer to this temperature as the bulk temperature of the maximally deformed BTZ theory.~\footnote{In
a Schwarzschild-like radial coordinate, $r\equiv(R^2-\rho^2)/R$, the maximally deformed BTZ metric,~(\ref{limit}), reads
$ds^2={R^2\over r_5^2}d\tau^2-f(r)R^2d\varphi^2+{r_5^2dr^2\over 4r^2f(r)}$ with $f(r)\equiv 1-{R\over r}$,
and the string coupling vanishes in the limit, $e^{\Phi(r)}\to 0$, for any $r\neq 0$.
Note that while the bulk temperature of the maximally deformed BTZ geometry is finite, $\beta_{bh}=2\pi r_5$,
the bulk mass, $M_{bh}$, and bulk entropy, $S_{bh}$, a.k.a. the mass and entropy of the $2d$ black hole with compact time in~$(\varphi,r)$,
are infinite, $S_{bh}=\beta_{bh}M_{bh}=2\pi e^{-2\Phi(\rm horizon)}=\infty$.
Needless to say that exposing the significance of these properties in negative single-trace $T\bar T$ holography is desired.}

Finally, the energy, entropy and temperature of the deformed BTZ theory in~(\ref{bs})--(\ref{vr5}),
$E$, $S$ and $T$ in~(\ref{e}),~(\ref{entropy}) and~(\ref{t}), respectively,
approach their maximal value,~(\ref{emax}),~(\ref{smax}) and (\ref{tmax}), respectively, in the limit.

Now, consider de Sitter black holes.

The Reissner-Nordstrom de Sitter metric is given by (see e.g. section 2 of~\cite{Montero:2019ekk})~\footnote{The
metric is supported by an electric (or magnetic) field, $A={Q\over r}dt$.}
\be\label{rnds}
ds^2=-f(r)dt^2+f(r)^{-1}dr^2+r^2d\Omega^2~,\qquad f(r)=1-{r^2\over \ell^2}-{2GM\over r}+{Q^2\over r^2}~.
\ee
The three positive roots,
\be\label{coi}
r_c\geq r_o\geq r_i~,
\ee
are the locations of the cosmological, outer and inner horizons of the black hole (BH) in de Sitter ($dS_4$), respectively.

For simplicity,~\footnote{And since the location of the inner horizon $r_i$ will not play a significant role below.}
we will focus on Schwarzschild de Sitter black holes (whose geometry is~(\ref{rnds}) with $Q=0$),
and extremal de~Sitter black holes (whose geometry is~(\ref{rnds}) with $Q=GM$), in turn.

Schwarzschild de Sitter black holes have
\be\label{ri}
r_i=0\qquad (Q=0)~.
\ee
In this case (see e.g.~\cite{Susskind:2021dfc}),
the location $r_{c,o}$ of the cosmic and event horizons and the de~Sitter size $\ell$ satisfy the relation
\be\label{lrr}
\ell^2-r_c^2-r_o^2={1\over 3}\left(\ell^2-(r_c-r_o)^2\right)~,\qquad 0\leq r_c-r_o\leq\ell~.
\ee
The total entropy $S$ of a black hole in de Sitter is the sum of the black hole and cosmic entropies,
\be\label{sdsbh}
S={\pi\over G}\left(r_c^2+r_o^2\right),
\ee
hence,~(\ref{lrr}),
\be\label{sds}
S=S_{dS}\left[1-{1\over 3}\left(1-\left({r_c-r_o\over\ell}\right)^2\right)\right]~,
\ee
where
\be\label{sdsempty}
S_{dS}={\pi\ell^2\over G}
\ee
is the entropy of empty de Sitter.

Extremal de Sitter black holes have
\be\label{rio}
r_i=r_o={\ell\over 2}\left(1-\sqrt{1-{4Q\over\ell}}\right)~,\quad r_c={\ell\over 2}\left(1+\sqrt{1-{4Q\over\ell}}\right)\qquad (Q=GM)~,
\ee
and hence entropy,~(\ref{sdsbh}),
\be\label{sext}
S=S_{dS}\left(1-{2Q\over\ell}\right)=S_{dS}\left[1-{1\over 2}\left(1-\left({r_c-r_o\over\ell}\right)^2\right)\right]~,
\ee
instead.

A few of comments are in order:
\begin{enumerate}
\item
The regime between the event horizon and cosmological horizon, $r_o<r<r_c$, is referred to as the static patch.~\footnote{The
static patch plays an important role in de Sitter holography; we shall return to this below.}
The regime beyond the cosmological horizon, $r>r_c$, is time dependent
(with time being $r$ instead of $t$ in the static patch).~\footnote{The
interior of the black hole, $r_i<r<r_o$, will not play a role below.}
\item
As one varies the location of the black hole horizon $r_o$ (while keeping the de~Sitter radius $\ell$ fixed),
one interpolates between empty de Sitter (when $r_o=0$) and Nariai geometry,
when the black hole horizon and the cosmic horizon coincide, $r_o=r_c$ (a.k.a. when the de~Sitter black hole size is maximal).
\item
The entropy is maximal,
\be\label{sdsmax}
S_{max}=S_{dS}~,
\ee
for empty de Sitter ($r_o=0$, in which case $r_c=\ell$,~(\ref{lrr}),~(\ref{rio})),
and is minimal when the de Sitter black hole size is maximal ($r_o=r_c$).~\footnote{The minimal value of the entropy is different
in the different branches: $S_{min}={2\over 3}S_{dS}$ in the Schwarzschild branch,~(\ref{sds}),
while $S_{min}={1\over 2}S_{dS}$ in the extremal branch,~(\ref{sext}), instead.}
\item
The inverse Hawking temperature of empty de Sitter is~\footnote{In de Sitter holography, it is
assumed though that the formal temperature in the Boltzmann distribution is infinite in empty de Sitter, $T_{Boltzmann}=\infty$,
while the finite Hawking temperature is a derived or emergent quantity,~\cite{Lin:2022nss}; we shall return to this point below.}
\be\label{bds}
\beta_{dS}=2\pi\ell~.
\ee
\end{enumerate}

The above suggests an analogy between the properties of de Sitter black holes and the negatively deformed BTZ black holes,
a.k.a. black holes in $\mathcal{M}_3^-$; it is summarized in table 1.~\footnote{The regimes that do not appear in table 1,
$r<r_o$ and $\rho<\rho_-$, do not have a manifest analogy.
This is due to the coordinate $\varphi$ of $\mathcal{M}_3^-$, on top of $(\tau,\rho)$,
playing an important role beyond the UV wall, at $\rho>R$,~\cite{Chakraborty:2020swe,Chakraborty:2020cgo}, on the one side,
and $dS_4$ ending at $r\to\infty$, on the other side;
correspondingly, the regime in the interior of the Reissner-Nordstrom de Sitter black hole (a.k.a. $r_i\leq r<r_o$)
and the regime beyond the inner horizon of the black hole in ${\cal M}_3^-$ (a.k.a. $0\leq\rho<\rho_-$), respectively,
do not participate in the analogy table.}

\begin{table}[h!]
  \begin{center}
    \label{tab:table1}
    \begin{tabular}{l|r}
     {\bf Black hole (BH) in $\mathcal{M}_3^- $} & {\bf Black hole (BH) in de Sitter ($dS_4$)} \\
      \hline
      Singular UV wall of $\mathcal{M}_3^-$ at $\rho=R$ & Outer horizon of BH in $dS_4$ at $r=r_o$\\
      \hline
     Outer horizon of BH in $\mathcal{M}_3^- $ at $\rho=\rho_+$ &  Cosmological horizon of $dS_4$ at $r=r_c$ \\
     \hline
     Inner horizon of BH in $\mathcal{M}_3^- $ at $\rho=\rho_-$ &  The end of the de Sitter world at $r\to\infty$ \\
     \hline
     Static patch of BH in $\mathcal{M}_3^- $ at $R>\rho>\rho_+$ & Static patch of BH in $dS_4$ at $r_o<r<r_c$ \\
     \hline
     Time-dependent regime at $\rho_+>\rho>\rho_-$ & Time-dependent regime at $r_c<r<\infty$ \\
     \hline
     Empty $\mathcal{M}_3^- $ (a.k.a. $\rho_+=0)$ & Largest BH in $dS_4$ (a.k.a. $r_o=r_c$)\\
     \hline
     $S_{min}$ when $\rho_+=0$ (a.k.a. empty $\mathcal{M}_3^- $) & $S_{min}$ when $r_o=r_c$ (a.k.a. largest BH)\\
     \hline
     Largest BH in $\mathcal{M}_3^- $ (a.k.a. $\rho_+=R)$ & Empty de Sitter (a.k.a. $r_o=0$)\\
     \hline
     $S_{max}$ when $\rho_+=R$ (a.k.a. largest BH) & $S_{max}$ when $r_o=0$ (a.k.a. empty $dS_4$)\\
     \hline
     $T\to\infty$ when $\rho_+\to R$  & $T_{Boltzmann}\to\infty$ when $r_o\to 0$\\
     \hline
     $\beta_{bh}=2\pi r_5$ for largest BH in $\mathcal{M}_3^- $ & $\beta_{dS}=2\pi\ell$ in empty de Sitter\\
    \end{tabular}
    \caption{The analogy table.}
  \end{center}
\end{table}

Consequently, there is an analogy between negative single-trace $T\bar T$ holography
(see e.g.~\cite{Giveon:2017nie,Chakraborty:2023mzc,Chakraborty:2023zdd} and references therein)
and de Sitter holography (see e.g.~\cite{Lin:2022nss,Chandrasekaran:2022cip,Susskind:2023rxm} and references therein),
which we discuss below.

In both de Sitter and negatively deformed BTZ geometries,
the holographic principle may be applied to the static patch in a similar~\footnote{Though opposite 
(e.g. the state of maximum finite entropy and infinite temperature amounts on the one hand to empty de Sitter, 
whose static patch size is maximal, $r_c-r_o\to\ell$,
and on the other hand to a maximally filled $\mathcal{M}_3^- $,
whose static patch size is minimal, $R-\rho_+\to 0$; see more details below).} spirit.
Next, we present the proposed principles, in turn.

In de Sitter holography, the proposal is the following
(see~\cite{Chandrasekaran:2022cip} and references therein for details).
Assume that the dimension of the Hilbert space that describes all possible states of the static patch,
including any matter and/or black holes it may contain and also those associated with its cosmological horizon,
is measured roughly by the area of the cosmological horizon of empty de Sitter, a.k.a. $e^{\pi\ell^2\over G}\equiv e^{S_{dS}}$,~(\ref{sdsempty}).
Now, place a particle with energy $E$ at the center of the static patch.
For e.g. a black hole with mass $M=E$,~(\ref{rnds}), this reduces the size of the cosmological horizon to
$r_c=\ell-GE+{\cal O}(E^2,Q^2)$, and so
\be\label{andso}
{\pi r_c^2\over G}=S_{dS}-\beta_{dS}E~,
\ee
to leading order in $E,Q$, where $\beta_{dS}$ is given in~(\ref{bds}).
Since the total number of microstates is $e^{S_{dS}}$, and the number of microstates corresponding to a particle with energy $E$
is $e^{A_c/4G}$, where $A_c=4\pi r_c^2$ is the area of the cosmological horizon, then assuming that there is an equal probability to observe
any particle with the same $E$ at the center of the static patch, the probability to observe such a particle is
\be\label{prob}
P(E)=e^{-\beta_{dS} E}~.
\ee
Consequently, the density matrix of the static patch of empty de~Sitter space is proportional to the identity,
a.k.a. empty de Sitter space is described by the maximally mixed state of the finite dimensional Hilbert space proposed above.

Equivalently, the formal temperature in the Boltzmann distribution is infinite,
while the finite Hawking temperature is a derived or emergent quantity.
Concretely, it is argued that a bulk can emerge at a finite effective temperature from a hologram at infinite temperature
(see~\cite{Lin:2022nss,Susskind:2023rxm} and references therein for details).

To recapitulate, the proposal is that empty de Sitter space maximizes the entropy of any state of the static patch,
$S_{max}=S_{dS}$,~(\ref{sdsmax}),
and consequently its horizon temperature is infinite, $T_{Boltzman}=\infty$, while its bulk is in an effective low temperature,
$T_{dS}={1\over 2\pi\ell}$,~(\ref{bds}).

In negative single-trace $T\bar T$ holography, the proposal is the following
(see e.g.~\cite{Giveon:2017nie,Chakraborty:2023mzc,Chakraborty:2023zdd} and references therein).

First, recall (see~\cite{Chakraborty:2023zdd} for details), that eqs.~(\ref{n}),~(\ref{entropy}), and~(\ref{e}),
can be rewritten as
\be\label{nr}
n={p\over\alpha'}\tilde\rho_-\tilde\rho_+~,
\ee
\be\label{sr}
S={2\pi kp\over r_5}\tilde\rho_+~,
\ee
\be\label{er}
E={p\over\lambda R}\left[-1+\sqrt{\left(1-{\tilde\rho_-^2\over R^2}\right)\left(1-{\tilde\rho_+^2\over R^2}\right)}\right]~,
\ee
respectively, where $\lambda<0$ is the one in~(\ref{l}), and
\be\label{trtr}
\tilde\rho_-\equiv{\rho_-\over\sqrt{1-{\rho_+^2\over R^2}}}~,\qquad
\tilde\rho_+\equiv{\rho_+\over\sqrt{1-{\rho_-^2\over R^2}}}~.
\ee
Consequently, on the $\lambda$-deformed BTZ backgrounds~(\ref{bs})--(\ref{vr5}) with fixed angular momentum $n$ and entropy $S$,
a.k.a. with fixed $\tilde\rho_\pm$, the $p$'th fraction of the energy $E(\lambda)$ takes the form
\be\label{el}
{1\over p}E(\lambda)={1\over\lambda R}\left(-1+\sqrt{1+2\lambda R{E(0)\over p}+\left({\lambda RP\over p}\right)^2}\right)~,
\ee
where
\be\label{mp}
E(0)\equiv{r_5\over R}M_{btz}~,\qquad P\equiv{n\over R}
\ee
are the energy and momentum of the states in the $CFT_2$ hologram of $AdS_3$ that amount to a BTZ black hole with mass $M_{btz}$
and angular momentum $n$.
Finally, in terms of the hologram quantities,
the $p$'th fraction of the entropy reads
\be\label{seplr}
{1\over p}S=2\pi\sqrt{c\over 6}\left(\sqrt{E_L(1+\lambda E_R)}+\sqrt{E_R(1+\lambda E_L)}\right)~,\qquad c\equiv 6k~,
\ee
where
\be\label{elr}
E_{L,R}\equiv{R\over 2p}\left(E(\lambda)\pm P\right)
\ee
are the $p$'th fractions of the left and right handed dimensionless energies, ${R\over 2}(E\pm P)$.

Equations~(\ref{el})--(\ref{elr}) are the same as what one would obtain by studying a symmetric product $CFT_2$,
${\cal M}^p/S_p$.~\footnote{On a circle with radius $R$,
and with the central charge of the block ${\cal M}$ being $c=6k$, as in~(\ref{seplr}).}
The generic state with large energy and momentum in the undeformed symmetric product CFT has (at large $E,P,p$)
its energy and momentum equally split among the different copies of ${\cal M}$.
Equation~(\ref{el}) gives the deformed energy of such states under a $T\bar T$ deformation of the seed CFT, ${\cal M}$,
with a dimensionless deformation parameter $\lambda$,~(\ref{l}), and eq.~(\ref{seplr}) is the corresponding entropy in the block.

To recapitulate, in single-trace $T\bar T$ holography, the proposal is thus that the physics of deformed BTZ black holes
is captured by the properties of a symmetric product theory whose block is a $T\bar T$ deformed CFT
(with the energy and momentum distributed equally in each of the blocks).~\footnote{The
status of the proposed symmetric product hologram is summarized in~\cite{Chakraborty:2023mzc} and references therein;
we shall not present it here.}
Indeed, one can check that there is a detailed match between the properties of the geometries~(\ref{bs})--(\ref{vr5})
and the corresponding energies and entropies,~(\ref{el})--(\ref{elr}), in the proposed holographic dual.
In particular, since the energy~(\ref{emax}) is a maximal energy in negatively deformed BTZ, for all $n$,
it is natural to conjecture that an analogous statement is true for standard $T\bar T$ deformed CFT with negative $\lambda$
(see~\cite{Chakraborty:2023zdd} and references therein for more details).
Consequently, the negatively deformed single-trace $T\bar T$ theory has a state of maximum finite entropy,~(\ref{smax}),
and this state is maximally mixed,~(\ref{tmax}).

In~(\ref{limit})--(\ref{givesc}), it was shown that the geometry which corresponds to the maximal energy and entropy,~(\ref{emax})
and~(\ref{smax}), respectively, and infinite temperature, is universal
(in the sense that it does not depend on the particular value of $n$); the static patch of the deformed BTZ geometries
is degenerated in these cases, $\rho_+\to R$, and the resulting background is an $SL(2)_k/U(1)$ black hole with compact time
(times a non-compact spatial direction).
In addition to $T=\infty$,~(\ref{t}),~(\ref{tmax}), there is thus a bulk temperature, a.k.a. that of the superstring theory on
$SL(2)_k/U(1)$, $T_{bh}={1\over 2\pi r_5}\left(={1\over 2\pi\sqrt{\alpha' k}}\right)$ in~(\ref{smax}) (with~(\ref{vr5})).
This appears in harmony with the proposed properties of de Sitter holography, presented above; see the analogy table 1.
There are however intriguing differences. In the following, we shall discuss both the similarities and differences.

In both negative single-trace $T\bar T$ holography and de Sitter holography,
the fact that there is a state of maximum finite entropy can be understood from the geometry.
In the former, it corresponds to the largest black hole in negatively deformed BTZ.
In the latter, it corresponds to the largest cosmological horizon size in de Sitter space, a.k.a. to empty de Sitter.

In de Sitter holography, it was argued~\cite{Lin:2022nss} that for the empty de Sitter state,
the horizon system, at $r=r_c=\ell$, is all that there is,
a.k.a. the bulk,~(\ref{rnds}) with $M=Q=0$, is not a second system; it is a holographic construct made of the horizon degrees of freedom.
In negative single-trace $T\bar T$ holography, for the maximal black hole state,
the static patch shrinks to the location of the event horizon, at $\rho=\rho_+=R$,
and one may say analogously that the horizon system is all that there is,
a.k.a. the bulk~(\ref{limit}) isn't a 2nd system; it's a holographic construct made of the horizon d.o.f.~\footnote{This is
similar to what is argued~\cite{McGough:2016lol} in the case of a geometric cutoff that removes the asymptotic region of $AdS_3$
and places the QFT on a Dirichlet wall at finite radial distance $r=R$ in the bulk; in our maximally deformed BTZ geometry,
the UV wall is the $\rho=\rho_+=R$ one.}.

In negative single-trace $T\bar T$ holography, we have a concrete proposal for the description of the microstates
via holography~\footnote{Apart for a measure zero amount of states (see~\cite{Chakraborty:2023mzc} and references therein).}
-- those of single-trace $T\bar T$ deformed symmetric product $CFT_2$ with maximal energy.
On the other hand, it seems fair to say that the description of the microstates
that constitute the entropy associated with cosmological de~Sitter horizons is more mysterious.

In both negative single-trace $T\bar T$ holography and de Sitter holography,
it is proposed that the state with maximum entropy is maximally mixed.
Equivalently, the temperature parameter in the Boltzmann distribution must be infinite, $T=\infty$.
In the former, such an infinite temperature is supported by explicit properties of both the geometry and the proposed holographic theory.
On the other hand, there is no concrete way to obtain this property from the de Sitter geometry,
and it seems fair to say that an explicit holographic candidate of de Sitter gravity with these properties is still a mystery.

As to the bulk inverse temperatures, $\beta_{dS}=2\pi\ell$ and $\beta_{bh}=2\pi r_5$ in empty de Sitter,~(\ref{bds}),
and maximally deformed BTZ,~(\ref{smax}), respectively, while they can be obtained from the surface gravity
both in empty de~Sitter and maximally deformed BTZ geometries, there is an educational difference that we discuss next.

The deformed BTZ geometry,~(\ref{bs})--(\ref{nn}), has an asymptotically linear dilaton regime beyond the naked singularity at $\rho=R$,
which plays an important role in negative single-trace $T\bar T$ holography
(see~\cite{Chakraborty:2020swe,Chakraborty:2020cgo} for details).
In particular, the signature of $\tau$ and $\varphi$ beyond the UV wall, $\rho>R$, is flipped relative to the static patch, $\rho_+<\rho<R$.
While the time in the holographic single-trace $T\bar T$ deformed symmetric product is that in the static patch of~(\ref{bs},\ref{nn}), $\tau$,
the timelike direction in the asymptotically linear dilaton regime is $\varphi$, instead, and the bulk $T_{bh}$ above is
w.r.t. to $\varphi$ being time,~(\ref{givesb}),~(\ref{givesc}), instead of~$\tau$.
This allowed us to obtain both $T\to\infty$ and $T_{bh}=1/2\pi r_5$ from the $\rho_+\to R$ limit of the
geometry~(\ref{bs},\ref{nn}).~\footnote{Recall~\cite{Chakraborty:2020swe,Chakraborty:2023zdd} that the temperature in eq.~(\ref{t}) is obtained
from the geometry e.g. by Wick rotating the canonically normalized~$t\equiv R\tau/r_5$, $t\to i\tau_E$,
compactifying $\tau_E$ on a circle with circumference $\beta=1/T$,
and demanding regularity of the metric~(\ref{bs},\ref{nn}) at the event horizon, $\rho=\rho_+$.
In particular, the $T\to\infty$ result is obtained from the $\rho_+\to R$ limit of the geometry in this way.
On the other hand, the bulk temperature, $T_{bh}=1/2\pi r_5$, is obtained from the $\rho_+\to R$ limit of the geometry
e.g. by Wick rotating the canonically normalized bulk time~$y\equiv R\varphi$ in eq.~(\ref{givesc}),
$y\to i\varphi_E$, compactifying $\varphi_E$ on a circle with circumference $\beta_{bh}=1/T_{bh}$,
and demanding regularity of the metric in~(\ref{givesc}) at the horizon, $\phi=0$, of the $SL(2)_k/U(1)$ black hole.}

It is important to recall though that obtaining the bulk inverse temperature in maximally deformed BTZ,
from the surface gravity at the horizon of an $SL(2)_k/U(1)$ black hole,
required us to extend the deformed BTZ geometry beyond its black hole singularity,~(\ref{rhophic}),~(\ref{givesc}).
This is not the case in de Sitter space, whose bulk temperature is obtained from the surface gravity at its cosmological horizon.

On the other hand, the de Sitter geometry does not have the analog of the asymptotically linear dilaton regime of~(\ref{bs}).
Consequently, the empty de Sitter geometry reveals only a single temperature -- the bulk $T_{dS}=1/2\pi\ell$ above,
while the proposed infinite Boltzmann temperature in the hologram is hidden from the geometry.

To recapitulate, the analogy and difference between negative single-trace $T\bar T$ holography and de~Sitter holography
is hinting towards the temptation to seek an extension  of the de Sitter geometry,
in a way that will give rise to an analog of the $\rho>R$ regime in the negatively deformed BTZ geometry~(\ref{bs}).

\vspace{10mm}

\section*{Acknowledgments}
I thank Soumangsu Chakraborty and David Kutasov for discussions.
This work was supported in part by the ISF (grant number 256/22) and the BSF (grant number 2018068).

\vspace{10mm}

\section*{Appendix -- A comment on $2$pf}

According to~\cite{Susskind:2023rxm}, there is a contradiction between the semiclassical fixed-background approximation and the holographic description in which the density matrix is maximally mixed, since in the former, the two point functions ($2$pf) have a non-zero imaginary part,
but in de~Sitter holography it must vanish.
This paradox and its proposed resolution within the principles of de Sitter holography are discussed in~\cite{Susskind:2023rxm}.

This is analogous to the fact that in ${\cal M}_3$ with an $R^{1,1}$ boundary,
while the $2$pf in the zero winding sector, $w=0$,
which amounts to the semiclassical fixed-background approximation (up to corrections in $1/k$,~(\ref{vr5})),
has a non-zero imaginary part,~\cite{Asrat:2017tzd},
in the sectors with non-zero winding, $w\neq 0$, a.k.a those that amount to stringy properties of the theory,
it doesn't,~\cite{Cui:2023jrb,Giveon:2023gzh}.

It is possible that in a full fledged consistent string theory on the negatively deformed BTZ black holes in~(\ref{bs})--(\ref{vr5}),
the states in the $w=0$ sector, with a non-zero imaginary part, are projected out, leaving behind only real two point functions.
The investigation of this issue is left for future work.~\footnote{For that purpose,
one should be able to extend the calculations in~\cite{Asrat:2017tzd,Cui:2023jrb,Giveon:2023gzh}
to the superstring theory on~(\ref{bs})--(\ref{vr5}),
in particular, for the $\rho_+\to R$ case,~(\ref{limit}).}


\begin{thebibliography}{1}

\bibitem{Chakraborty:2020swe}
S.~Chakraborty, A.~Giveon and D.~Kutasov,
``$ T\overline{T} $, black holes and negative strings,''
JHEP \textbf{09}, 057 (2020)
[arXiv:2006.13249 [hep-th]].

\bibitem{Chakraborty:2023mzc}
S.~Chakraborty, A.~Giveon and D.~Kutasov,
``Comments on single-trace $ T\overline{T} $ holography,''
JHEP \textbf{06}, 018 (2023)
[arXiv:2303.12422 [hep-th]].

\bibitem{Chakraborty:2023zdd}
S.~Chakraborty, A.~Giveon and D.~Kutasov,
``Momentum in Single-trace $T\bar T$ Holography,''
[arXiv:2304.09212 [hep-th]].

\bibitem{Giveon:2005mi}
A.~Giveon, D.~Kutasov, E.~Rabinovici and A.~Sever,
``Phases of quantum gravity in AdS(3) and linear dilaton backgrounds,''
Nucl. Phys. B \textbf{719}, 3-34 (2005)
[arXiv:hep-th/0503121 [hep-th]].

\bibitem{Giveon:2017nie}
A.~Giveon, N.~Itzhaki and D.~Kutasov,
``$ \mathrm{T}\overline{\mathrm{T}} $ and LST,''
JHEP \textbf{07}, 122 (2017)
[arXiv:1701.05576 [hep-th]].

\bibitem{Giveon:1994fu}
A.~Giveon, M.~Porrati and E.~Rabinovici,
``Target space duality in string theory,''
Phys. Rept. \textbf{244}, 77-202 (1994)
[arXiv:hep-th/9401139 [hep-th]].

\bibitem{Montero:2019ekk}
M.~Montero, T.~Van Riet and G.~Venken,
``Festina Lente: EFT Constraints from Charged Black Hole Evaporation in de Sitter,''
JHEP \textbf{01}, 039 (2020)
[arXiv:1910.01648 [hep-th]].

\bibitem{Susskind:2021dfc}
L.~Susskind,
``Black Holes Hint towards De Sitter Matrix Theory,''
Universe \textbf{9}, no.8, 368 (2023)
[arXiv:2109.01322 [hep-th]].

\bibitem{Lin:2022nss}
H.~Lin and L.~Susskind,
``Infinite Temperature's Not So Hot,''
[arXiv:2206.01083 [hep-th]].

\bibitem{Chakraborty:2020cgo}
S.~Chakraborty, A.~Giveon and D.~Kutasov,
``Strings in irrelevant deformations of AdS$_{3}$/CFT$_{2}$,''
JHEP \textbf{11}, 057 (2020)
[arXiv:2009.03929 [hep-th]].

\bibitem{Chandrasekaran:2022cip}
V.~Chandrasekaran, R.~Longo, G.~Penington and E.~Witten,
``An algebra of observables for de Sitter space,''
JHEP \textbf{02}, 082 (2023)
[arXiv:2206.10780 [hep-th]].

\bibitem{Susskind:2023rxm}
L.~Susskind,
``A Paradox and its Resolution Illustrate Principles of de Sitter Holography,''
[arXiv:2304.00589 [hep-th]].

\bibitem{McGough:2016lol}
L.~McGough, M.~Mezei and H.~Verlinde,
``Moving the CFT into the bulk with $ T\overline{T} $,''
JHEP \textbf{04}, 010 (2018)
[arXiv:1611.03470 [hep-th]].

\bibitem{Asrat:2017tzd}
M.~Asrat, A.~Giveon, N.~Itzhaki and D.~Kutasov,
``Holography Beyond AdS,''
Nucl. Phys. B \textbf{932}, 241-253 (2018)
[arXiv:1711.02690 [hep-th]].

\bibitem{Cui:2023jrb}
W.~Cui, H.~Shu, W.~Song and J.~Wang,
``Correlation Functions in the TsT/$T{\bar T}$ Correspondence,''
[arXiv:2304.04684 [hep-th]].

\bibitem{Giveon:2023gzh}
A.~Giveon,
``2pf in single-trace $ T\overline{T} $ holography,''
JHEP \textbf{10}, 112 (2023)
[arXiv:2309.15629 [hep-th]].


\end{thebibliography}
\end{document}